\documentclass[a4paper,11pt]{article}
\usepackage{amssymb}
\usepackage{amsmath}
\usepackage{graphicx}

\def\circa#1{\,\raise.3ex\hbox{$#1$\kern-.75em\lower1ex\hbox{$\sim$}}\,}

\font\tenrsfs=rsfs10 at 11pt
\font\sevenrsfs=rsfs7
\font\fiversfs=rsfs5
\newfam\rsfsfam
\textfont\rsfsfam=\tenrsfs
\scriptfont\rsfsfam=\sevenrsfs
\scriptscriptfont\rsfsfam=\fiversfs
\def\mathscr#1{{\fam\rsfsfam\relax#1}}

\makeatletter

\newcounter{alphaequation}[equation]
\def\thealphaequation{\theequation\hbox to
0.6em{\hfil\alph{alphaequation}\hfil}}
\def\eqnsystem#1{
\def\@eqnnum{{\rm (\thealphaequation)}}
\def\@@eqncr{\let\@tempa\relax \ifcase\@eqcnt \def\@tempa{& & &} \or
  \def\@tempa{& &}\or \def\@tempa{&}\fi\@tempa
  \if@eqnsw\@eqnnum\refstepcounter{alphaequation}\fi
\global\@eqnswtrue\global\@eqcnt=0\cr}
\refstepcounter{equation} \let\@currentlabel\theequation \def\@tempb{#1}
\ifx\@tempb\empty\else\label{#1}\fi
\refstepcounter{alphaequation}
\let\@currentlabel\thealphaequation
\global\@eqnswtrue\global\@eqcnt=0 \tabskip\@centering\let\\=\@eqncr
$$\halign to \displaywidth\bgroup \@eqnsel\hskip\@centering
$\displaystyle\tabskip\z@{##}$&\global\@eqcnt\@ne
\hskip2\arraycolsep\hfil${##}$\hfil& \global\@eqcnt\tw@\hskip2\arraycolsep
$\displaystyle\tabskip\z@{##}$\hfil
\tabskip\@centering&\llap{##}\tabskip\z@\cr}

\def\endeqnsystem{\@@eqncr\egroup$$\global\@ignoretrue} \makeatother

\def\pl#1#2#3{Phys.~Lett.~{\bf B {#1}} ({#2}) #3}

\begin{document}
\tolerance=100000
\thispagestyle{empty}
\setcounter{page}{0}
~
\vspace{3cm}

\begin{center}
{\LARGE \bf 
Majorana Dark Matter.
}\\[1.5cm]

{\large\bf Alexey Anisimov}\\[3mm]
{\it Institut de theorie des phenomenes physiques,\\
Ecole polytechnique federale de Lausanne,\\
Lausanne, 1015, Switzerland} \\[10mm]

{\large\bf Abstract}
\end{center}
\begin{quote}
{\noindent
In this letter in the framework of a simple see-saw scenario with two (or three) quasi degenerate Majorana neutrinos we propose\footnote{This letter is based on the original work \cite{paper}.} that one of these neutrinos can be very weakly coupled, yet there is a mechanism of the generation of the abundance of such "dark" neutrino in the early universe. The mechanism of production is due to oscillations between "dark" Majorana neutrino and one of the Majorana's which has relatively large Yukawa couplings ("bright" Majorana neutrino). The transition of "bright" Majorana into a "dark" one is nonadiabatic. We point out on the similarity with the Landau-Zener transition regime. In our model one can explain the observed dark matter density, present matter-antimatter asymmetry and active neutrino data all at the same time. }

\end{quote}

\newpage

\setcounter{page}{1}

\section{Introduction}

The observed mater-antimatter asymmetry and the presence of yet unknown form of matter  in the universe known as Dark Matter (DM) both require the extension of the Standard Model which does not contain neither a DM candidate nor a mechanism to produce large enough baryon asymmetry in order to match the observed value. On the other hand, the observed neutrino mass scales and mixings can not be incorporated within the Standard Model (SM) as well. The sew-saw mechanism \cite{seesaw} provide probably the most natural and simple explanation why neutrinos are massive, yet so light compared to all other Standard Model particles. At the same time, neutrino mixing data represent a positive test for leptogenesis \cite{fy}, an attractive explanation of the the observed matter-antimatter asymmetry of the Universe as a direct consequence of the see-saw mechanism. Can DM be accommodated within the see-saw model as well?  It is known, that it is enough to have just two Majorana's to account for both leptogenesis and active neutrino data (see, e.g. \cite{glashow}). Assuming that the number of Majorana neutrinos is the same as the number of lepton flavors in Standard Model one can ask the question if one of the Majorana neutrino can be made almost decoupled and, thus, stable at present but being produced efficiently enough in the early universe.  However, if a particle is almost stable at present there is typically quite hard to produce it during earlier epoch. 
One clear possibility is the DM production from the inflaton decay along with the rest of particles. There is no argument why not to assume, for example, that inflaton couplings to the DM particle and to the SM matter are not of the same order. In that case DM particle can be produced as efficiently as ordinary matter. However, such mechanism, due to the present lack of the detailed picture of inflation and most importantly the exit from it, is strongly model dependent. Besides, nothing would require one of the Majorana neutrinos to be a DM particle. Any other particle decoupled from the SM fields would be as plausible. It is appealing, therefore, to look for another more constrained mechanism of the production of the "dark" Majorana neutrino. The perturbative production out of the thermal bath can be a solution. However, typically this would require the reheating temperature after inflation to be quite high.   
 
In this letter we propose that "dark" Majorana neutrino is produced {\it via} oscillation mechanism which is similar to the Landau-Zener type transition. The production starts and terminates near some critical temperature $T_c$ due to the oscillations between "dark" Majorana neutrino and one of the strongly coupled (or "bright") Majorana neutrino. The critical temperature can be as low as the Majorana neutrino mass scale which, in turn, can be as low as ${\mathcal O}({\rm TeV})$. Contrary to the perturbative production mechanism in our model high reheating temperature is not required. In order for oscillation mechanism to be successful both neutrinos have to be quasi degenerate in mass.\footnote{In the case of quasi degenerate neutrinos the Majorana mass scale can be made quite low as the consequence of resonant enhancement of the kinematic part of the expression for the CP-asymmetry, which compensates for the smallness of the Yukawa couplings \cite{res}.} We will discuss two scenarios. The one, called "minimal", has only the see-saw type couplings being added to the SM Lagrangian. In this case the production rate turns out to be not efficient enough. In the "next-to-minimal" scenario we will introduce a simple dimension five  operator which couples Higgs field to Majorana neutrinos and is suppressed by the high energy physics  scale $\Lambda$. Such operator could result from GUT extensions of the SM, supergravity etc. This operator is nearly irrelevant for any physics at present. However, at relatively high temperatures ($T>T_{\rm EW}$) it will turn out to be an effective source for the "dark" Majorana production.\footnote{A simultaneous picture of baryogenesis and DM has been already proposed \cite{asaka,asaka2}. Here
the matter-antimatter asymmetry is generated through RH neutrino oscillations \cite{smirnov} and the lightest
${\cal O}$(KeV) mass RH neutrino is a warm DM. The other two Majorana neutrinos in this scenario are much heavier but still have to be lighter then the electroweak scale as well. Previously, some other models were already proposed where the light sterile neutrino is produced by the mixing with the ordinary light ones and plays the role of Dark Matter \cite{warm}. In this letter we explore the case when Majorana neutrinos mass scale is well above the electroweak scale.}  

\section{The Model.}

The see-saw extension of the Standard Model is described by the Lagrangian
\begin{equation}
\begin{array}{c}
{\mathcal
L}={\mathcal
L}_{\rm SM}+i{\bar
N}_i{\hat\partial}N_i-F^{ik}{\bar
L}_iN_k\Phi^c-{M_i\over 2}{\bar
N}_i^c N_i+{\rm h.c}.
\end{array}
\label{lagr}
\end{equation} 
For simplicity we will assume that only one of the "bright" Majorana neutrino serves as a source for the production of the "dark" Majorana neutrino. We will employ the density matrix formalism \cite{dolgov,siglraffelt}. The evolution of the number densities of the Majorana and of the active neutrinos is described by the density matrix which obeys quantum kinetic equation:
\begin{equation}
i{\dot\rho_{\bf k}}=[H_{\bf k},\rho_{\bf k}]-{i\over 2}\{\Gamma^d_{\bf k}+\Gamma^p_{\bf k},\rho\}+i\Gamma^p_{\bf k},
\label{basic}
\end{equation}
where $\Gamma^{d,p}_{\bf k}$ are destruction and production rates correspondingly (see, e.g. \cite{weldon}).
In general case, this is $2\times(3+N_M)$ matrix equation, where $N_M$ is the number of the Majorana neutrinos. In practice, one can always neglect the off-diagonal elements $\rho_{NL}$ of this matrix  because the latter undergo rapid oscillations.
Because the helicity flip and lepton number violation processes are suppressed the opposite helicity parts of this matrix evolve independently. This simplifies the Eq.~(\ref{basic}) further. Finally, in our model the oscillations of "bright" neutrino into a "dark" one is effective only during a brief period, so that light lepton part of the density matrix does not have any effect on the final abundance of the "dark" Majorana neutrino.  Therefore, it is enough to consider only  the part of this matrix which describes two quasi degenerate Majorana neutrinos: one "dark" and one "bright". This is described by  $2\times 2$ matrix the $\rho=(\rho)_{AB}$ is  ($A,B$ are flavor indices). 
This  equation, in general, has to be solved for each mode ${\bf k}$ and the resulting spectral matrix $\rho_{\bf k}$ then has to be integrated over the phase volume to obtain the number density matrix $\rho$:
\begin{equation}
\rho=\int{d^3{\bf k}\over(2\pi)^3}\rho_{\bf k}.
\label{rho}
\end{equation}
The diagonal elements of the matrix $\rho$ give the physical number densities of the two neutrinos after being multiplied by the massless fermion number density (or, equivalently, by the photon number density $n_{\gamma}$).

In our model one of the neutrinos is decoupled and the other is in thermal equilibrium.\footnote{We assume that the critical temperature $T_c$ satisfies the condition $M<T_c<T^{\rm eq}$, where $M$ is the Majorana neutrino mass scale and $T^{\rm eq}$ is the temperature at which the second neutrino come to the thermal equilibrium. This condition put some constraints on the model, that will be discussed later.} In that case the Eq.~(\ref{basic}) simplifies to 
\begin{equation}
i{\dot\rho_{\bf k}}=[H_{\bf k},\rho_{\bf k}].
\label{basic1}
\end{equation}
Furthermore, instead of solving the Eq.~(\ref{basic1}) for each mode we will adopt a more simplified approach choosing $k\sim T$ in order to facilitate further discussion. A more quantitative analysis will be given elsewhere \cite{paper}. In this letter we aim on a qualitative discussion so that the latter simplification is justified. Because the abundance of the "dark" Majorana neutrino is expected to be much smaller then thermal abundance of the "bright" Majorana neutrino at the moment of production, we can set $\rho_{BB}=\rho^{\rm eq}_{BB}$. In addition we will assume that there is no CP violation in the $A-B$ sector. With all these assumptions the Eq.~(\ref{basic1}) for a given momentum   $k\sim T$ leads to the following set of equations:
\begin{eqnarray}
\nonumber
\dot\rho_{AA}&=&~2H_{AB}\rho_-,\\
\nonumber
\dot\rho_-~~&=&\rho_+(H_{BB}-H_{AA})-H_{AB}\rho_{BB}^{\rm eq},\\
\nonumber
\dot\rho_+~~&=&\rho_-(H_{AA}-H_{BB}),\label{diff_set}
\end{eqnarray}
where we have defined $\rho_{+}=(\rho_{AB}+\rho_{BA})/2$,
$\rho_{-}=-i\,(\rho_{BA}-\rho_{AB})/2$.
The Hamiltonian can be computed from the real part of the Majorana neutrino two-point Green function on Fig.~(1). Below we will consider two cases with two different effective temperature dependent Hamiltonians result from self-energy corrections
shown on Fig.~(2-3).
\begin{figure}
\begin{center}
\includegraphics[width=4.7in]{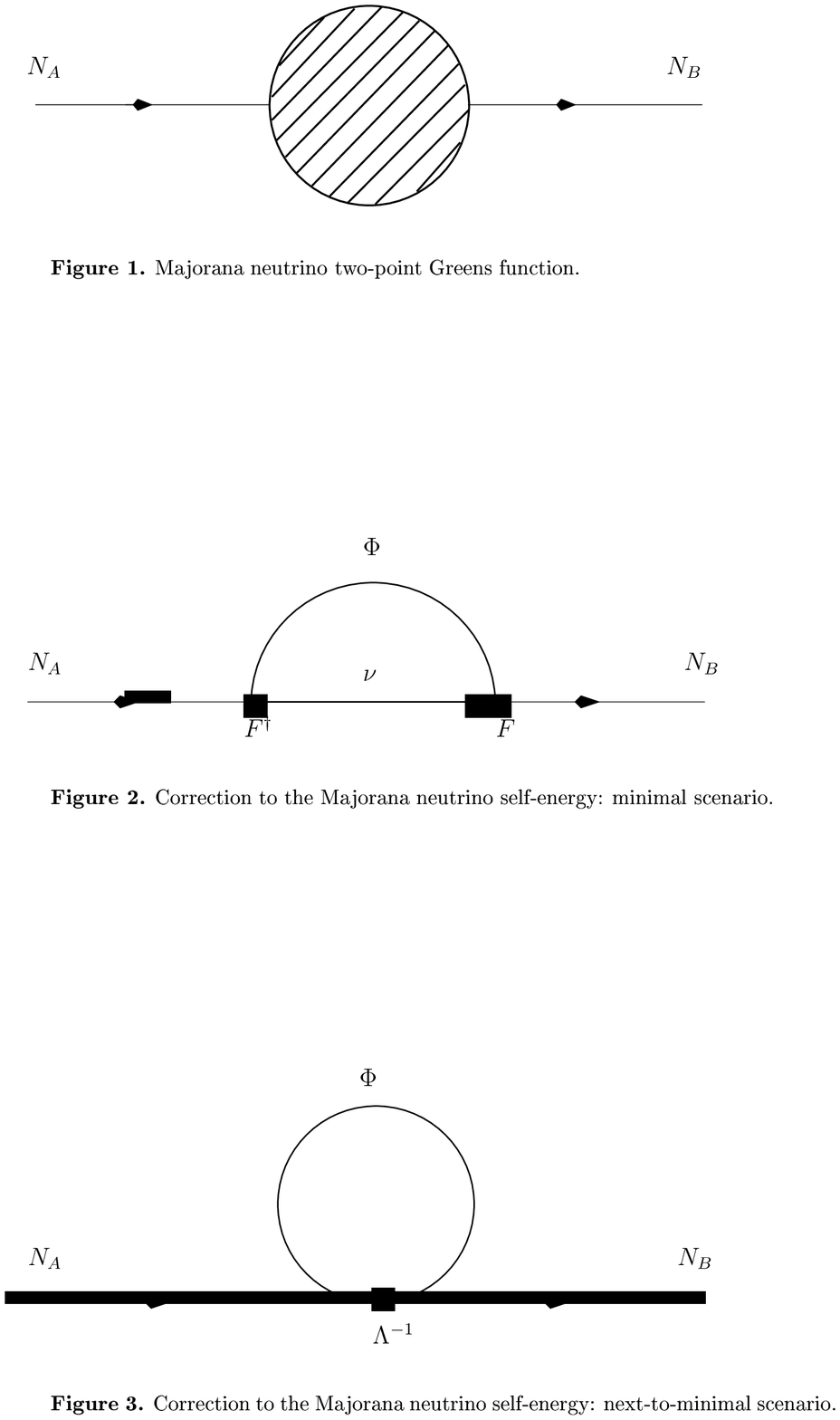}
\end{center}
\end{figure}

\section{The Minimal Scenario.}

The Lagrangian of the model (\ref{lagr}) is written in the Majorana neutrino mass eigenstate basis. The Yukawa matrix in the case of two generations is given by\footnote{The indices $i,j$ will be used for the mass eigenstate basis, while letter indices will represent flavor basis. The index $A$ is always reserved for the "dark" Majorana neutrino.}
\begin{equation}
(F)_{ij}= \left(
\begin{array}{cc} F_{11} & F_{12} \\
F_{21} & F_{22} \\
\end{array} \right)\approx \left(
\begin{array}{cc} 0 & F_{12} \\
F_{21} & F_{22} \\
\end{array} \right),
\end{equation}
where by assumption  $F_{11}\ll F_{12,21}\ll F_{22}$, therefore we neglect $F_{11}$ keeping only $F_{12,21}$ elements.
The Hamiltonian can be computed following \cite{weldon1} and is given by\footnote{Assuming that the two RH neutrinos are strongly degenerate,
such that $\Delta\equiv |M_2-M_1|/M_1 \ll 1$.}
\begin{equation}
(H)_{ij}=\sqrt{M_1^2+k^2}\,I+
{T\over 8}\,\left(\begin{array}{cc}
K_{11} & K_{12} \\
K_{21} & K_{22}+{8\,\Delta M_{21}^2\over T\,\sqrt{M_1^2+k^2}},
\end{array}\right)
\end{equation}
where $I$ is the identity matrix and the matrix $K$ is defined as
\begin{equation}
K=F^{\dagger}F.
\end{equation}
The first term, proportional to the identity matrix,
cancels out in the evolution equation so that we will not keep it further.
The computation of the "dark" Majorana neutrino production can be made in the mass eigenstate basis by computing
$\rho_{11}$ and noting that $\rho_{11}\approx\rho_{AA}$. Instead, we will make a rotation into the flavor basis and 
compute $\rho_{AA}$ directly. The diagonalization of the matrix $F_{ij}$ can be made with the help of two unitary 
matrices $U$ (rotation of the Majorana neutrino basis) and $V$ (rotation of the light lepton basis). Introducing two different angles $\theta_U$ and $\theta_V$ and diagonalizing matrix $F_{ij}$ one obtains
\begin{equation}
(F)_{AB}\approx \left(
\begin{array}{cc} -{F_{12}F_{21}\over F_{22}} & 0 \\
0 & F_{22} \\
\end{array} \right),
\end{equation}
where it was used that both angles $\theta_{U,V}$ are small and given by
\begin{equation}
\theta_{U,V}\approx {F_{21,12}\over F_{22}}.
\end{equation}
The Hamiltonian is rotated according to
\begin{equation}
H'= U^{\dagger}HU,
\end{equation}
and one obtains the Hamiltonian  familiar from the active neutrino oscillations and MSW effect:
\begin{equation}
(H')_{AB}\simeq\left(
\begin{array}{cc} 0 &  {M^2\theta_U\Delta\over E}\\
  {M^2\theta_U\Delta\over E} & -{M^2\Delta\over E}+{1\over 8}TK_{22}\\
\end{array} \right),
\end{equation}
where $E=\sqrt{M^2+k^2}$, $M=M_1\simeq M_2$.
Assuming that $E\approx k\sim T$, $\Delta M^2_{21}\equiv-2M^2\Delta<0$, where $\Delta\equiv(M_1-M_2)/M_2\ll 1$  
one obtains solving the Eqs.~(\ref{diff_set}) 
\begin{equation}
\rho_{AA}\approx {m_{pl}\over  M}\left(K_{22}\over 8\Delta\right)^{1\over 2}
K_{11}\le {m_{pl}\over 2 M}K_{11},
\label{rho_AA}
\end{equation}
where $m_{pl}=M_{pl}\sqrt{90/8\pi^3g_*}\sim 10^{18}{\rm GeV}$.
The last inequality follows from the requirement that the critical temperature $T_c=M(8\Delta/K_{22})^{1/2}$ 
at which $H'_{BB}$ becomes equal to zero satisfies $T_c>M$.

It is instructive to compare (\ref{rho_AA}) to the value of the Landau-Zener parameter $Q$ that controls the regime in which the transition occurs. Its value is given by (see, e.g. \cite{kim})
\begin{equation}
Q={M^2\Delta\sin^22\theta_U\over E\cos 2\theta_U h_R},
\label{Q}
\end{equation}
where 
\begin{equation}
h_R=\left|{2\dot T\over T}\right|={2T^2\over m_{pl}}.
\end{equation}
For MSW effect \cite{MSW} it is essential that $Q\gg 1$ while in the Landau-Zener regime $Q\le 1$.  
Taking into account account that $\theta_U\ll 1$ and estimating for $k\sim T$ the value of (\ref{Q}) 
one obtains at $T=T_c$
\begin{equation}
Q={m_{pl}\over M}\left(K_{22}\over 8\Delta\right)^{1\over 2}
K_{11}.
\end{equation}
One can see that $\rho_{11}\approx Q$, as expected since both represent the probability
of the transition in the nonadiabatic regime.
The upper bound on the DM energy density can be estimated as follows
\begin{equation}
{\Omega_{DM} h^2\over 0.1}\approx 10^{25}K_{11}.
\label{omega}
\end{equation}
Given that the lifetime of the flavor state $N_A$ has to exceed the age of the universe in order to be a DM candidate one can estimate that $K_{11}\le 10^{-42}(1~{\rm TeV}/M_1)$ and the corresponding density (\ref{omega}) is by far too small.

\section{The next-to-minimal scenario.}

As it was shown in the previous section the minimal model fails to produce the correct amount of the ''dark'' Majorana neutrino abundance. In this section we propose an extension to the previous setup by introducing (explicitly) no new particles but allowing a dimension five effective operator 
\begin{equation}
{\mathcal L}_{\rm eff}={(\lambda)_{AB}\over \Lambda}|\Phi|^2\bar N_A^cN_B,
\label{eff}
\end{equation}
where $\Phi$ is usual Higgs field, $\Lambda$ is some high energy scale and the elements of the matrix $(\lambda)_{AB}\sim{\mathcal O}(1)$.
The correction of the type (\ref{eff}) is generic if one assumes a certain additional physics at energies much higher then the electroweak scale. For example the operator (\ref{eff}) generically will result from supergravity with $\Lambda$ being $M_{pl}$ or from tadpole type graphs in GUT theories  
{\it via} heavy Higgs exchange. There are other possibilities to generate (\ref{eff}). In what follows we will not address any specific model which provides the relation of the scale $\Lambda$ to any other scales like $M_{pl},M_{\rm GUT}, M_{\rm SUSY}$ etc. Instead we will treat $\Lambda$ as phenomenological parameter and impose the constraints on its value from the requirement that the operator (\ref{eff}) is responsible for the correct DM abundance at present.
Additionally, we assume that
\begin{equation}
{F_{12}\Delta M\over F_{22}}\ll {T_{\rm EW}^2\over \Lambda}.
\label{c1}
\end{equation}
The latter conditions simply implies that the the operator  (\ref{eff}) in the Majorana neutrino flavor basis gives a contribution to the off-diagonal entrees of the Hamiltonian which dominate over the term 
\begin{equation}
{\Delta M^2\sin2\theta\over 2E}
\label{common}
\end{equation}
at all temperature above the electroweak phase transition. In that case we may disregard (\ref{common}) in the computation below. The validity (\ref{c1}) can be easily inforced by choosing $\theta_U$ small enough (or even setting it to zero).

The resummed Majorana propagator will contain the nondiagonal entrees which are given by
\begin{equation}
-\Sigma_{AB}\approx{1\over\Lambda}\int^{\infty}_{0}{d^4q\over (2\pi)^3}\delta(q^2-m^2_{\Phi})n_{\rm b}(q),
\end{equation}
where 
\begin{equation}
n_{\rm b}(q)={1\over e^{|q\cdot u|\over T}-1},
\end{equation}
with $u$ being a four velocity of the thermal bath. The resulting expression for the real finite part of $\Sigma_{AB}$ in Boltzmann approximation is
\begin{equation}
-{\rm Re}[\Sigma_{AB}-\Sigma_{AB}(T=0)]\approx{T^2\over 2\pi^2\Lambda}z_{\Phi}K_1(z_{\Phi})u_{\mu}\gamma^{\mu},
\end{equation}
where $z_{\Phi}=(m_{\Phi}/T)\sim {\mathcal O}(1)$, and $K_1(z)$ is a modified Bessel function. Alternatively, we can neglect thermal mass of the Higgs. In that case one obtains
\begin{equation}
-{\rm Re}[\Sigma_{AB}-\Sigma_{AB}(T=0)]\approx{T^2\over 12\Lambda}u_{\mu}\gamma^{\mu}.
\label{m0}
\end{equation}
In what follows the exact coefficient in front of $T^2/\Lambda$, which is of order of ${\mathcal O}(10^{-1})$ is not particularly important and we will use (\ref{m0}) for simplicity.
Assuming as well that\footnote{This assumption is not crucial but simplifies analytics and turns out to be satisfied for a wide range of parameters after restrictions coming from the "dark" neutrino abundance at present and its decay rate are imposed.} 
\begin{equation}
{1\over 8}K_{22}\gg {T_c\over \Lambda}.
\label{c2}
\end{equation}
results in the following effective Hamiltonian:
\begin{equation}
H\approx \left(
\begin{array}{cc} 0 &  -{T^2\over 12\Lambda}\\
 -{T^2\over 12\Lambda} & -{M^2\Delta\over E_{\bf k}}+{1\over 8}TK_{22}\\ 
\end{array}\right).
\end{equation}
Solving Eq.~(\ref{basic1}) one obtains that
\begin{equation}
\rho_{AA}\sim {m_{pl}M\over z_c\Lambda^2K_{22}},
\label{int}
\end{equation}
where $z_c$ is given by 
\begin{equation}
z_c=\left({K_{22}\over 8\Delta}\right)^{1\over 2}.
\end{equation}
The above estimate  transfers into the DM abundance which is at present 
\begin{equation}
{\Omega_{DM}h^2\over 0.1}\approx {1\over z_c}\left({10^{24}{\rm GeV}\over \Lambda}\right)^2\left(M\over 
10^{9}{\rm GeV}\right).
\label{dm1}
\end{equation}
The operator (\ref{eff}) after electroweak symmetry breaking leads to the mixing term $(v_{\rm EW}^2/\Lambda)\bar{N}_AN_B$ between "dark" and "bright"
neutrinos. This results in the decay of the "dark" neutrino {\it via} oscillations into the "bright" one and its consequent
decay. The rate of this process is
\begin{equation}
\Gamma(N_A\to N_B\to ...)=\left(v_{\rm EW}^2\over\Lambda\right)^2{\Gamma_B\over\left(\Gamma_B\over 2\right)^2+M^2\Delta^2}.
\end{equation}
The decay rate of the "bright" neutrino is dominated by $\Gamma(N_B\to\Phi+\nu)$ and one estimate that
\begin{equation}
\Gamma(N_A\to N_B\to ...)\sim 10^{-35}z_c^4\left(10^{24}{\rm GeV}\over \Lambda\right)^2\left(10^9{\rm GeV}\over 
M\right)^2{\rm eV}.
\end{equation}
There is phase space of parameters $\{M,\Lambda,z_c\}$ which leads to the correct DM abundance and the lifetime of the "dark" neutrino larger then the age of the Universe. For example, one can choose $z_c=10^{-2}$, $M=10^7~{\rm GeV}$ and $\Lambda=10^{24}{\rm GeV}$. For this choice of parameters the condition (\ref{c2}) is satisfied as well. The "extreme" case would be to take $z_c=10^{-3}$ (because, at smaller values it is hard to have the "bright" neutrino being thermally produced), $M=1~{\rm TeV}$ and $\Lambda\sim 10^{22.5}{\rm GeV}$. In this case the "dark" neutrino has the mass near electroweak scale and a lifetime close to the present age of the Universe. Of course, a more stringent bounds may come from astrophysical observations and the lower bound on the neutrino mass scale quoted here may go up substantially. 

There is an another constraint which comes from the four-body decay $N_A\to 2\Phi+N_B\to 3\Phi+\nu$.
The rate of such decay can be estimated as
\begin{equation}
\Gamma_4\sim 10^{-7}{M^3\over \Lambda^2}K_{22}\sim 10^{-24}\left({M\over 10^{9}{\rm GeV}}\right)^4\left({10^{24}{\rm GeV}\over \Lambda}\right)^2{\rm eV}.
\end{equation}
Requiring that this rate is smaller then the value of the Hubble parameter at present leads to the constraint
\begin{equation}
\left({M\over 10^{9}{\rm GeV}}\right)^2\left({10^{24}{\rm GeV}\over \Lambda}\right)\le 10^{-4.5}.
\end{equation}
This implies that $M$ can not be higher then about $10^{8}-10^{9}{\rm GeV}$ depending on the value of $z_c$.

\section{Conclusions}

In this letter we proposed the see-saw model  in which one of three Majorana neutrinos ("dark" neutrino) is weakly coupled and is a DM candidate. The mechanism of the generation of a suitable abundance was suggested. At least one of the strongly coupled Majorana neutrino ("bright" neutrino) have to be quasi degenerate with the "dark" one. The allowed quasi degenerate Majoranas mass scale $M$ can, in principle, be as low as ${\mathcal O}({\rm TeV})$ and up to as high as usually suggested in GUT scenarios, i.e. $M\sim 10^8-10^{9}{\rm GeV}$. The mass of the third neutrino is adjusted correspondingly: if $M$ is below $10^{8}{\rm GeV}$ the third neutrino has to be also quasi degenerate with the other two in order to have lepton number produced efficiently. Otherwise it remains a relatively free parameter as in the standard hierarchical leptogenesis scenario. This letter is based on the results which will be reported in much more details in \cite{paper}.


\end{document}